\documentstyle[11pt]{article}
\makeatletter
\def\section{\@startsection{section}{1}{\z@}{-3.5ex plus -1ex minus -2.ex}
{2.3ex plus .2ex}{\Large\bf}}

\def\subsection{\@startsection{subsection}{2}{\z@}{-3.25ex plus
 -1ex minus -2.ex}
{1.5ex plus .2ex}{\bf}}

\def\vsn{\vskip 1pc \noindent}
\def\f{\newline}
\def\e{\varepsilon}

\def\comp{{\rm comp}}

\def\rr{{\bf R}}

\def\P{{\rm{\bf P}}}
\def\E{{\rm{\bf E}}}

\newcommand{\be} {\begin{equation}}
\newcommand{\ee} {\end{equation}}
\newcommand{\bd} {\begin{displaymath}}
\newcommand{\ed} {\end{displaymath}}

\newcommand{\bq}{\begin{eqnarray}}
\newcommand{\eq}{\end{eqnarray}}
\newcommand{\bqn}{\begin{eqnarray*}}
\newcommand{\eqn}{\end{eqnarray*}}
\newcommand{\ba}[1]{\begin{array}{#1}}
\newcommand{\eqa}{\end{array}}
\def\qed{
   \\[-4ex]
  \hbox to \hsize{\hfill \vrule height 1.6ex width 1.5ex
  depth -.1ex}}

\begin{document}

\bibliographystyle{alpha}

\begin{center} {\Large {\bf
Improved  Bounds on the Randomized and Quantum Complexity  of
Initial-Value Problems
 }\footnotemark[1] }
 
\end{center}
\footnotetext[1]{ ~\noindent This research was partly supported by  
  AGH grant No. 10.420.03 \vsn}               
  
\medskip
\begin{center}
{\large {\bf Boles\l aw Kacewicz \footnotemark[2] }}
\end{center}
\footnotetext[2]{ 
\begin{minipage}[t]{16cm} 
 \noindent
{\it Department of Applied Mathematics, AGH University of Science 
and Technology,\\
\noindent  Al. Mickiewicza 30, paw. A3/A4, III p., 
pok. 301,\\
 30-059 Cracow, Poland 
\newline
 kacewicz@uci.agh.edu.pl, tel. +48(12)617 3996, fax +48(12)617 3165 }  
\end{minipage} }

\thispagestyle{empty}

\begin{center} {\bf{\Large Abstract}} \end{center}
{\small 
We study the problem, initiated  
in \cite{rand5}, of finding randomized and quantum complexity of
initial-value problems. We showed in \cite{rand5} that a~speed-up  in 
both settings over the 
worst-case deterministic complexity is possible.  In the present paper
we prove, by defining  new algorithms, that further 
improvement in upper bounds on the randomized and quantum 
complexity can be achieved. 
In the H\"older class of right-hand side functions with $r$ continuous 
bounded partial derivatives, with $r$-th derivative being
a H\"older function with exponent $\rho$, 
the $\e$-complexity is shown to be  
$O\left( (1/\e)^{1/(r+\rho+1/3)} \right)$ in the randomized setting, 
and $O\left( (1/\e)^{1/(r+\rho+1/2)} \right)$ on a quantum computer 
(up to logarithmic factors).
This is an improvement for the general problem over the results 
from \cite{rand5}. The gap still remaining between upper and 
lower bounds on the complexity is further discussed for a special problem.
We consider scalar autonomous problems, with the aim of computing 
the solution at the end point of the interval of integration.
For this problem, we 
fill up the gap by establishing (essentially) matching
upper and lower complexity bounds.  We show that the complexity in this case 
is 
$\Theta\left( (1/\e)^{1/(r+\rho+1/2)} \right)$ in the randomized 
setting, and
$\Theta\left( (1/\e)^{1/(r+\rho+1)} \right)$ in the~quantum setting
(again up to logarithmic factors).
Hence, this problem is essentially as hard as the~integration 
problem.
}
\newpage

\setcounter{page}{1}
\normalsize
\pagestyle{plain}
{\Large \section{ Introduction }}
\noindent
\vsn
Significant progress has been made in recent years in the field of quantum
complexity of numerical problems. Integration (\cite{Novak}, followed by
\cite{Heinrich}) was the first problem to be so studied. Other problems
were next analyzed, such as approximation \cite{Heinrich1,Heinrich2}
and path integration \cite{Wozn}. The only paper that has studied 
the randomized and quantum complexity of initial-value problems for 
ordinary differential equations is \cite{rand5}. This paper showed that 
we can achieve a nontrivial speed-up by going from the worst-case 
deterministic setting to the randomized or quantum settings. 
The~idea in \cite{rand5} was to use the optimal deterministic algorithm
based on integral information \cite{Kac1}, and replace integrals 
in a suitable way by optimal randomized or
quantum approximations \cite{N88,Novak}. We recall the results 
from \cite{rand5} in Theorem~1.
\f
In the present paper, we show that further improvement in upper bounds 
on the randomized and quantum complexity is possible. 
We first define a~new deterministic integral algorithm for 
initial-value problems (Section 3).  
Although this algorithm is not optimal in the deterministic 
worst-case setting, it is better suited for randomization and 
implementation on a quantum computer than the algorithm used in \cite{rand5}. 
Randomized and quantum algorithms are defined by a suitable application of 
optimal randomized and quantum algorithms for summation of real 
numbers \cite{Bras,Nayak} $\;$ (Section 4). 
The reduction of the total cost is achieved due to  
a better balance, compared to the algorithms from \cite{rand5},
between the deterministic and random components of the~cost. 
\f
New upper bounds on the complexity are shown  in Theorem~2 in Section 5. 
In the H\"older class of right-hand side functions with $r$ continuous 
bounded partial derivatives, with $r$-th derivative being
a H\"older function with exponent $\rho$, 
the $\e$-complexity is shown to be (up to logarithmic 
factors) $O\left( (1/\e)^{1/(r+\rho+1/3)} \right)$ in the randomized setting, 
and $O\left( (1/\e)^{1/(r+\rho+1/2)} \right)$ on a quantum computer.
Noticeable improvement in
both settings is thus achieved, compared to the bounds from Theorem~1. 
The gap between upper and lower complexity bounds is 
reduced (but still not cancelled). 
\f
In order to further reduce the gap between the bounds, we turn to  
a special case of the general problem. We study in Section 6 
the complexity of computing the~solution  of a scalar autonomous problem  
at one single point. 
In \cite{rand5}, we only showed (non-optimal) upper bounds on the 
randomized and quantum complexity of this problem. The question about 
lower bounds was left open. 
\f
We provide essentially matching upper and lower complexity bounds 
in Theorem~3. 
Upper bounds are established by using a bisection argument, while 
lower bounds by reducing the~problem to the summation of real numbers. 
Up to logarithmic factors, the complexity turns out to be 
$\Theta\left( (1/\e)^{1/(r+\rho+1/2)} \right)$ in the randomized 
setting, and
$\Theta\left( (1/\e)^{1/(r+\rho+1)} \right)$ in the quantum setting.
The gap between  upper and lower bounds 
is thus essentially closed. Up to logarithmic factors,
the problem considered turns out to be  as difficult as 
the integration problem. 
\vsn
{\Large \section{ Preliminaries  }}
\noindent
We deal  with the randomized and quantum solution of a system of ordinary 
differential equations with initial conditions
\be
z'(t)=f(z(t)), \;\;\; t\in [a,b],\;\;\;\;\; z(a)=\eta,
\label{1}
\ee
where $f:\rr^d \to \rr^d$, the initial vector $\eta$ is in $\rr^d$,  
and the~solution $z$ maps $[a,b]$ into $\rr^d$.
We assume that $f(\eta)\ne 0$. 
\f
This formulation covers nonautonomous systems 
$z'(t)=f(t,z(t))$ with  $f: \rr^{d+1} \to \rr^d$, which 
can be written in the form (\ref{1}) by adding one scalar equation:
$$ \left[ \begin{array}{l} u'(t) \\ z'(t)\end{array} \right]
= \left[ \begin{array}{l} 1 \\ f(u(t), z(t)) \end{array} \right]
$$
with an additional initial condition $u(a)=a$.
We assume that the right-hand side function $f=[f^1,\ldots,f^d]^T$ 
belongs to the H\"older class $F^{r,\rho}$.
Given an integer $r\geq 0$, a number $\rho\in (0,1]$, 
positive numbers $D_0,D_1,\ldots, D_r$ and $H$, we set 
$$
F^{r,\rho} =\{\, f:\rr^d\to \rr^d \mid \; f\in C^r(\rr^d), \;\;\;
|\partial ^i f^j(y)| \leq D_i, \; i=0,1,\ldots, r,
$$
\be
 |\partial^r f^j(y)-\partial^r f^j(z)|\leq 
H\, \|y-z\|^{\rho}, \; y,z \in \rr^d,\;\; j=1,2,\ldots, d \, \},
\label{2}
\ee
where $\partial^i f^j$ represents all partial derivatives of order $i$ 
of the $j$-th component of $f$,
and $\|\cdot\|$ denotes the maximum norm in $\rr^d$. 
We assume that $\rho=1$ for $r=0$, which assures that $f$ is 
a~Lipschitz function. 
\f
We formulate the problem and shortly recall  
basic definitions concerning randomized and quantum settings.
Our aim is to compute a bounded function $l$ on $[a,b]$ that approximates 
the~solution $z$. Letting $\{x_i\}$ be the uniform partition of $[a,b]$, 
so that $x_i=a+ih$ with $h=(b-a)/n$, we will construct  $l$ based on 
approximations 
$a_i(f)$ to $z(x_i)$ for $i=0,1,\ldots,n$. We assume that available information 
about the right-hand side $f$ is given by a subroutine that computes values 
of a component of $f$ 
or its partial derivatives. In the randomized setting, we allow for 
a random selection of points at which the values are computed. 
On a quantum computer, by subroutine calls we mean 
applications of a quantum query operator for (a component of) $f$, or
evaluations of  components of $f$ or its partial derivatives on a classical
computer. 
The transformation $\phi$ that computes $l$ based on available information is 
called an algorithm. 
\f
To be more specific, let  ($\Omega$, $\Sigma$, $\P$)
be a probability space. 
Let the mappings $\omega\in \Omega 
\mapsto  a_i^\omega(f)$
be random variables for each $f\in F^{r,\rho}$.
By an algorithm we mean a tuple
\be 
\phi =( \{a_0^\omega(\cdot),a_1^\omega(\cdot), \ldots, a_n^\omega(\cdot)
\}_{\omega\in \Omega},\psi),
\label{3.1}
\ee
where $\psi$ is a mapping that produces a bounded function $l^{\omega}$
based on $a_0^\omega(f)$, $a_1^\omega(f)$, $\ldots, a_n^\omega(f)$, 
\be
l^\omega(t)=\psi(a_0^\omega(f), a_1^\omega(f),\ldots, 
a_n^\omega(f))(t)\, ,
\label{3.2}
\ee
for $t\in [a,b]$. 
The error of $\phi$ at $f$ is defined by 
\be
e^\omega(\phi,f)=\sup_{t\in [a,b]}\|z(t)-l^\omega(t)\|.
\label{3.3}
\ee
We assume that the mapping
$\omega\in \Omega \rightarrow e^\omega(\phi,f)$
is a random variable for each $f\in F^{r,\rho}$.
\f
In the randomized setting, the error of $\phi$ in the class $F^{r,\rho}$ is
given by the maximal dispersion of $e^\omega(\phi,f)$,
\be
e^{{\rm rand}}(\phi, F^{r,\rho}) = \sup\limits_{f\in F^{r,\rho}}
({\rm {\bf E}} e^\omega(\phi,f)^2)^{1/2} \, ,
\label{3a}
\ee
where {\bf E} is the expectation.  (We could consider as well 
the maximal expected value of $e^\omega(\phi,f)$; this would only 
change the constants in our results.) 
The cost of an algorithm $\phi$ in the randomized setting is measured by a 
number of subroutine calls  needed to compute an approximation. 
For a given $\e>0$, by the $\e$-complexity of the problem, 
$\comp^{{\rm rand}} (F^{r,\rho}, \e)$, 
we mean the minimal cost of an algorithm $\phi$ taken among all $\phi$
such that  $e^{{\rm rand}}(\phi, F^{r,\rho}) \leq \e$.
\vsn
On a quantum computer, the output of an algorithm 
is also a random variable (taking a finite number of values). 
The randomness in the quantum setting results from quantum measurement 
operations \cite{Heinrich}. 
The right-hand side function $f$ 
can be accessed through applications of a quantum query operator $Q_f$
on a quantum space (defined through values of components of $f$).
Evaluations of components of $f$ or its partial derivatives
on a classical computer are also allowed.
For a~detailed discussion of the quantum query operator, and of the effect 
of quantum measurement, the reader is referred to \cite{Heinrich}. 
The error of an algorithm $\phi$ at $f$ in the quantum setting 
is again given by (\ref{3.3}), and 
the error of $\phi$ in the class $F^{r,\rho}$ by
\be
e^{{\rm quant}}(\phi, F^{r,\rho},\delta) = \sup\limits_{f\in F^{r,\rho}}
\inf\; \{\; \alpha|\;\; \P\{\, e^{\omega}(\phi,f)>\alpha \,\}\; 
\leq \delta\; \},
\label{3b}
\ee
for a given number $\delta$, where $0<\delta<1/2$.
For $\e>0$, (\ref{3b}) implies that the bound $e^\omega(\phi,f)\leq \e$ 
holds with probability at least $1-\delta$ for each $f$ iff
$e^{{\rm quant}}(\phi, F^{r,\rho},\delta) \leq \e$. Hence, $1-\delta$ 
is the~(minimal) success probability in computing an $\e$-approximation.
\f
The value of $\delta$ is usually set to $\delta=1/4$.
The success probability can then be increased to be at least $1-\delta$ 
(for arbitrarily small $\delta$) by computing component by component the 
median of $c\log 1/\delta$ repetitions of the algorithm, where $c$ 
is a positive number independent of $\delta$, see \cite{Heinrich1}. 
\f
The~cost of an algorithm $\phi$ in the quantum setting is measured by the
number of quantum queries, together with the number of classical 
evaluations of $f$ or its partial derivatives, needed to compute 
an approximation. 
For a given $\e>0$, by the quantum $\e$-complexity of the problem, 
$\comp^{{\rm quant}}(F^{r,\rho},\e,\delta)$, we mean 
the minimal cost of a quantum algorithm $\phi$ taken among all $\phi$
such that $e^{{\rm quant}}(\phi, F^{r,\rho},\delta) \leq \e$ . 
\vsn
We now recall upper and lower bounds on the randomized 
and quantum complexity for problem~(\ref{1}) 
obtained in \cite{rand5}. (We write below $\log$ for $\log_2$, 
although the base of the logarithm is not crucial.)
\vsn
{\bf Theorem 1}$\;\;$ (\cite{rand5}) $\;\;$ {\it For problem (\ref{1}), 
we have that
\f
\be
\comp^{{\rm rand}} (F^{r,\rho}, \e) = 
O\left( \left(\frac{1}{\e}\right)^
{ \frac{r+\rho +3/2}{(r+\rho +1/2)(r+\rho+1)} } 
\log \frac{1}{\e} \right) \, , 
\label{7}
\ee
\be
 \comp^{{\rm quant}} (F^{r,\rho}, \e, \delta) = 
O\left( \left( \frac{1}{\e}\right) ^{\frac{r+\rho+2}{(r+\rho+1)^2}} 
\left(\log \frac{1}{\e} + \log \frac{1}{\delta} \right) \right) \, . 
\label{6}
\ee
Moreover, for  $d\geq 2$
\be
\comp^{{\rm rand}} (F^{r,\rho}, \e) = 
\Omega \left( \left(\frac{1}{\e}\right)^{\frac{1}{r +\rho +1/2} } \right) \, ,
\label{7a}
\ee
and, for $0< \delta\leq 1/4$, 
\be
\comp^{{\rm quant}} (F^{r,\rho}, \e, \delta) \geq 
\comp^{{\rm quant}} (F^{r,\rho}, \e, 1/4) = 
\Omega \left( \left(\frac{1}{\e}\right)^
{\frac{1}{r+\rho+1}} \right) \, .
\label{6a}
\ee
The constants in the $O$- and $\Omega$-notation 
only depend on the class $F^{r,\rho}$, and are independent 
of $\e$ and $\delta$. 
} \qed
\vsn
In the deterministic worst-case setting, if only the values of $f$ or its 
partial derivatives can be accessed, the complexity of 
problem~(\ref{1}) is  $\Theta(\e^{-1/(r+\rho)})$. 
Hence,  Theorem~1 shows a~speed-up in both randomized and quantum settings
over the deterministic setting for all $r$ 
and $\rho$. Note also that there is a gap in the randomized and quantum  
settings   between the upper and lower complexity bounds given in Theorem~1. 
\f
In this paper, we show that further 
improvement in upper bounds on the randomized and quantum complexities
 is possible (Theorem~2). We start in the next section by defining 
 a~new deterministic algorithm that will be used to design  
 randomized and quantum algorithms in~Section~4. 
\f
In the next sections we shall need results on randomized and quantum 
computation of the mean of real numbers, which we now recall. 
Suppose we wish to compute the value 
\be
S=\frac{1}{s}\sum\limits_{i=1}^s x_i,
\label{7b}
\ee
for $-1\leq x_i\leq 1$. The $\e$-complexity of this problem in the 
randomized setting is defined as the~minimal number of 
accesses to $x_1, \ldots, x_s$ that is sufficient to find a random 
approximation 
$A^{\omega}$ to $S$ with expected error at most $\e$, 
$\E|A^{\omega} -S|\leq \e$. It is proportional to
\be
\min\{s, (1/\e)^2 \} 
\label{7c}
\ee
due to the result of Math\'e, see for a discussion \cite{HN}. 
Note that  $\E|A^{\omega} -S|\leq \e$ implies that 
\be
{\bf P}\{ |A^{\omega} - S| > 4\e \} \leq 1/4.
\label{71c}
\ee
On a quantum computer we can do better than this. 
The probabilistic error criterion (\ref{71c}) is used in the quantum setting, 
and the cost of an algorithm is measured
by a number of quantum queries (quantum accesses to $x_1,\ldots,x_s$). 
It is shown in \cite{Bras}
(upper bound) and \cite{Nayak} (lower bound) that the quantum complexity 
of computing the mean is proportional to
\be
\min\{s, 1/\e \}.
\label{7d}
\ee
\vsn
{\Large\section{Deterministic Algorithm 
}}
\noindent
We define a deterministic integral algorithm for solving (\ref{1}), which
will be the subject to randomization and implementation on 
a quantum computer in the next section. 
\f
Let $m,n\geq 1$. Define $\{x_i\}$ to be $n+1$ equidistant partition 
points of $[a,b]$, so that
$x_i=a+ih$ for $i=0,1,\ldots, n$, where $h=(b-a)/n$. 
Let $\{z_j^i\}$ define a partition of each interval 
$[x_i,x_{i+1}]$ with $m+1$ equidistant points $z_j^i= x_i + j \bar{h}$
for $ j=0,1,\ldots, m, \,$ with $\bar{h}=(x_{i+1}-x_i)/m$. 
Let $y_0^*=\eta$.  By induction, we define sequences
$\{y_i^*\}$ and $\{y_j^i\}$ as follows. For a given $y_i^*$ we set $y_0^i=y_i^*$.
Given $y_j^i$, by $z_{ij}^* $ we denote the solution of the local problem
\be
z_{ij} '(t)= f(z_{ij}(t)), \;\; t\in [z_j^i, z_{j+1}^i],\;\;\;\;
z_{ij}(z_j^i) = y_j^i . 
\label{p1}
\ee
Letting $l_{ij}^*(t)$ be defined by 
$l_{ij}^*(t)=\sum\limits_{k=0}^{r+1} (1/k!) 
z^{*\,(k)}_{ij}(z_j^i)(t-z_j^i)^k$ for $t\in [z_j^i,z_{j+1}^i]$, we set
$y_{j+1}^i = l^*_{ij}( z_{j+1}^i)$ for $j=0,1,\ldots, m-1$. 
Finally, we define
the function $l_i^*$ in $[x_i,x_{i+1}]$ by $l_i^*(t)= l_{ij}^*(t)$ for 
$t\in [z_j^i, z_{j+1}^i]$, and we compute the approximation to $z(x_{i+1})$
by
\be
y_{i+1}^* = y_i^* + \int\limits_{x_i}^{x_{i+1}} f(l_i^*(t))\, dt\, 
\qquad (0\leq i\leq n-1).
\label{p2}
\ee
The approximation~$l$ to the solution~$z$ of (\ref{1}) 
in $[a,b]$ is defined by 
\be
l(t)= l_i^*(t) \;\;\; \mbox{ for } t\in [x_i, x_{i+1}].
\label{p3}
\ee
Compared to the algorithm used in \cite{rand5}, 
the construction above is based not only on the points
$\{x_i\}$, but also on the finer partition given by $\{z_j^i\}$.
The approximation $l_i^*$ in $[x_i,x_{i+1}]$ is computed by  successive 
applications of Taylor's method with step size $\bar{h}$.
\f
In the sequel, we shall need an error bound for $l_i^*$ in 
$[x_i,x_{i+1}]$. The following lemma, stated without proof, 
is a standard result for Taylor's method, showing the dependence of the error 
on the length of the interval of integration.
Let $\bar{z}_i^*$ be the solution of the problem
\be
\bar{z}'(t)= f(\bar{z}(t)), \;\;\; t\in [x_i, x_{i+1}], \;\;\;\; 
\bar{z}(x_i)= y_i^*.
\label{p4}
\ee
{\bf Lemma}$\;\;\;$ {\it There exists a constant $M$ depending 
only on the parameters of the class $F^{r,\rho}$ 
(and independent of $i$, $y_i^*$ and $n$) such that
$$ \sup_{t\in [x_i, x_{i+1}]} \| \bar{z}_i^*(t) - l_i^*(t)\| \leq  
M\, h \bar{h} ^{r+\rho},$$
for sufficiently small $h$ ($Lh \leq \ln 2$, where $L$ is 
a~Lipschitz constant for $f$). 
} \qed
\vsn
The algorithm defined above is not optimal in the deterministic worst-case 
setting. It follows from this Lemma and the results from \cite{Kac1} that its
worst-case error in $[a,b]$ in the class $F^{r,\rho}$
is $O(1/(n(nm)^{r+\rho} ))$. This  
is achieved by using $\Theta(nm)$ evaluations. With the same number of 
evaluations it is however possible to get error $O(1/(nm)^{r+\rho +1})$, 
see \cite{Kac1}.
\vsn
In order to define randomized and quantum algorithms, we express (\ref{p2})
in an equivalent form. Defining
\be
w_{ij}^*(y) = \sum\limits_{k=0}^r \frac{1}{k!} f^{(k)}(y_j^i)(y-y_j^i)^k
\label{p5}
\ee
and 
\be
g_{ij}(u)= \frac{1}{\bar{h}^{r+\rho}} \left( f(l_{ij}^*(z_j^i+ u \bar{h})) -
w_{ij}^*(l_{ij}^* (z_j^i+ u \bar{h})) \right)\, , \;\;\;\; u\in [0,1]  ,
\label{p6}
\ee
we can write (\ref{p2}) as
\be
y_{i+1}^* = y_i^* + \sum\limits_{j=0}^{m-1}
\int\limits_{z_j^i}^{z_{j+1}^i} w_{ij}^* (l_{ij}^*(t))\, dt +
\bar{h}^{r+\rho+1} \sum\limits_{j=0}^{m-1}
\int\limits_{0}^{1} g_{ij} (u)\, du \, .
\label{p7}
\ee
Arguments similar to those used in the proof of Lemma in \cite{rand5} yield
(after replacing
the interval $[x_i,x_{i+1}]$ by $[z_j^i, z_{j+1}^i]$, $\, h$ by $\bar{h}$
and $y_i^*, l_i^*, w_i^*$ by $y_j^i, l_{ij}^*, w_{ij}^*$, respectively)
that the functions $g_{ij}$ are in $C^{(r)}([0,1])$, and the derivatives
of $g_{ij}$ of order $0,1,\ldots, r$ are bounded by constants depending
only on the parameters of the class $F^{r,\rho}$. Moreover, 
$$\|g_{ij}^{(r)} (u) - g_{ij}^{(r)} (\bar{u})\| \leq \tilde{H} 
|u-\bar{u}|^{\rho}, \;\;\; u,\bar{u} \in [0,1],  $$
where $\bar{H}$ is a constant depending only on the parameters of 
$F^{r,\rho}$.
{\Large\section{ Randomized and Quantum Algorithms}}
\noindent
We shall denote approximations obtained in randomized and quantum algorithms
by the same symbols as we did in the deterministic algorithm, 
omitting only the asterisk. In particular,
the~approximation to $z(x_i)$ is denoted by $y_i$. We start with $y_0=\eta$. 
For a given $y_i$ we put $y_0^i=y_i$, and denote by $z_{ij}$ the solution
of (\ref{p1}) (with the initial value $y_j^i$ computed for $y_i$).
We compute $l_{ij}$
in a same way as $l_{ij}^*$ (with $y_i$ instead of $y_i^*$), and we set
$y_{j+1}^i = l_{ij}(z_{j+1}^i)$.   Approximations 
$l_i$ in $[x_i,x_{i+1}]$ are defined to be equal to $l_{ij}$ in each 
subinterval $[z_j^i, z_{j+1}^i]$, 
and the polynomial $w_{ij}$ is constructed 
in the same way as $w_{ij}^*$, with $y_i^*$ replaced by $y_i$. 
\f
The approximation at $x_{i+1}$ is defined by
\be
y_{i+1}=y_i +\sum\limits_{j=0}^{m-1} \int\limits_{z_j^i}^{z_{j+1}^i}
w_{ij} (l_{ij}(t))\, dt + m \bar{h}^{r+\rho+1} A_i(f),
\label{p8}
\ee
where $A_i(f)$ is a  randomized or quantum approximation 
\be
A_i(f)\approx \frac{1}{m} \sum\limits_{j=0}^{m-1} 
\int\limits_0^1 g_{ij}(u)\, du\, .
\label{p9}
\ee 
The approximation $l$ in $[a,b]$ is defined by $l(t)=l_i(t)$
for $t\in [x_i,x_{i+1}]$. 
\f
For comparison, in \cite{rand5} we had $m=1$ and $A_i(f)$ was 
taken to be 
optimal randomized or quantum approximation to the integral 
$\int\limits_0^1 g_{i0}(u)\, du$.
\f
Here, we define $A_i(f)$ in a different way. Let $Q_{ij}^N(f)$ be 
the mid-point rule 
approximation to $\int\limits_0^1 g_{ij}(u)\, du$ based on $N$ points,
\be
Q_{ij}^N (f) = \frac{1}{N} \sum\limits_{k=0}^{N-1} g_{ij}(u_k)\, .
\label{p10}
\ee
Consider the first-stage approximation (without computing it) 
\be
\frac{1}{m} \sum\limits_{j=0}^{m-1} 
\int\limits_0^1 g_{ij}(u)\, du \approx 
\frac{1}{m} \sum\limits_{j=0}^{m-1} Q_{ij}^N(f) = 
\frac{1}{mN} \sum\limits_{j=0}^{m-1} \sum\limits_{k=0}^{N-1} g_{ij}(u_k).
\label{p11}
\ee
We define $A_i(f)$ to be the optimal randomized or
quantum approximation (computed component by component) 
to the right-hand side mean of $mN$ vectors in (\ref{p11}). 
\f
Consider first the quantum setting. Let $\e_1 >0$. 
For $i=0,1,\ldots, n-1$, let $A_i(f)$ be a random variable such that 
\be
\P\biggl\{ \biggl\|  A_i(f) - \frac{1}{mN} \sum\limits_{j=0}^{m-1} 
\sum\limits_{k=0}^{N-1} g_{ij}(u_k) \biggr\| \leq \e_1 \biggr\} 
\geq \frac{3}{4}
\label{p12}
\ee
for all $f\in F^{r,\rho}$. To compute $A_i(f)$ it suffices to use
 of  order $\min\{mN, 1/\e_1\}$  quantum queries 
 for computing each component of the mean, see (\ref{7d}). 
 (A number of repetitions 
dependent on~$d$ is also needed to keep the success probability 
at least $3/4$ when passing from components to the vector norm. This 
changes the cost by a constant factor only.)
To increase the success probability, we take 
the median (computed component by component) of $k$ results $A_i(f)$,
where 
$$k =\Theta \left(\log 
 \frac{1}{ 1- (1-\delta)^{1/n} } \right) =O( \log\, n +\log \, 1/\delta)$$
(with absolute constants in the $\Theta$- and $O$-notation). 
We get a~new approximation, denoted by the same symbol $A_i(f)$,  such that
\be
\P\biggl\{ \biggl\|  A_i(f) - \frac{1}{mN} \sum\limits_{j=0}^{m-1} 
\sum\limits_{k=0}^{N-1} g_{ij}(u_k) \biggr\| \leq \e_1 \biggr\} \geq 
(1-\delta)^{1/n}.
\label{p13}
\ee
This yields that 
\be
\P\biggl\{ \biggl\| A_i(f) - \frac{1}{mN} \sum\limits_{j=0}^{m-1} 
\sum\limits_{k=0}^{N-1} g_{ij}(u_k) \biggr\| \leq \e_1\; \mbox { for }
i=0,1,\ldots, n-1 \biggr\} \geq 1-\delta .
\label{p14}
\ee
The cost of computing $A_i(f)$ is  
$O\left( n  (\log\, n +\log\, 1/\delta) \min \{mN, 1/\e_1\} \right)$ 
quantum queries.
\f
In the randomized setting, we compute  each component of the mean
using the algorithm with expected error at most $\e_1/4$, and 
cost proportional to $\min\{mN, (1/\e_1)^2 \}$, see (\ref{7c}). 
Inequality (\ref{71c}) then holds with $\e:=\e_1/4$. We next proceed  
as in the quantum case above to compute $A_i(f)$ such that  (\ref{p14}) holds. 
For this, we need $O\left( n \, (\log\, n +\log\, 1/\delta) 
\min \{mN, (1/\e_1)^2 \} \right)$ function evaluations. 
\vsn
The deterministic part of the cost of algorithm (\ref{p8}) consists
of computing coefficients of $l_{ij}$ and $w_{ij}$ for $j=0,1,\ldots, m-1$,
for which we need $c m$ evaluations of partial derivatives of $f$ of
order $0,1,\ldots, r$, where $c$ only depends on $r$ and $d$. 
The computation of the integrals of $w_{ij}$ does not require new evaluations. 
Taking into account all indices $i$ and $j$,  we need in total $c n m$ 
evaluations of $f$ or its  partial derivatives.
\f
{\Large\section{ Upper Bounds on the Randomized and Quantum Complexity}}
\noindent
We now prove new upper bounds on the complexity of (\ref{1}).
\vsn
{\bf Theorem 2}$\;\;$ {\it For problem (\ref{1}), there exist constants
$P_1$ and $P_2$ depending only on the parameters of the class $F^{r,\rho}$
such that for sufficiently small $\e$ and $\delta$, 
\f
\be
\comp^{{\rm rand}} (F^{r,\rho}, \e) \leq
P_1 \left(\frac{1}{\e}\right)^{ 1/(r+\rho +1/3) } \log \frac{1}{\e} 
\label{th21}
\ee
and 
\be
\comp^{{\rm quant}} (F^{r,\rho}, \e, \delta) \leq
P_2 \left( \frac{1}{\e}\right) ^{1/(r+\rho+1/2)} 
\left( \log \frac{1}{\e} + \log \frac{1}{\delta} \right)  \, . 
\label{th22}
\ee }
\f
{\bf Proof}$\;\;$ 
We analyze the error of the algorithm defined in the previous section. 
Let $e_i=z(x_i)-y_i$. Since
\be
z(x_{i+1})=z(x_i) + \sum\limits_{j=0}^{m-1} \int\limits_{z_j^i}^{z_{j+1}^i} 
f(z(t))\, dt\, ,
\label{p15}
\ee
by subtracting (\ref{p8}) we get that
\be
e_{i+1} = e_i +  \sum\limits_{j=0}^{m-1} \int\limits_{z_j^i}^{z_{j+1}^i} 
(f(z(t)) - f(l_{ij}(t))) \, dt +
\sum\limits_{j=0}^{m-1} \int\limits_{z_j^i}^{z_{j+1}^i} 
(f(l_{ij}(t)) - w_{ij}(l_{ij}(t))) \, dt - m\bar{h}^{r+\rho+1} A_i(f).
\label{p16}
\ee
Hence,
\be
\|e_{i+1}\| \leq  \|e_i\| +  \sum\limits_{j=0}^{m-1} 
\int\limits_{z_j^i}^{z_{j+1}^i} 
\|f(z(t)) - f(l_{ij}(t)) \| \, dt +
m\bar{h}^{r+\rho+1}
\biggl{\|} \frac{1}{m} \sum\limits_{j=0}^{m-1} \int\limits_{0}^{1} 
g_{ij}(u) \, du -  A_i(f)\biggr{\|}
\label{p17}
\ee
for $i=0,1,\ldots, n-1$, where the function $g_{ij}$ is defined for $y_i$.
\f
Let $\bar{z}_i$ be the solution of (\ref{p4}) with the initial 
condition $\bar{z}(x_i)= y_i$.  Using the well known dependence of 
the solution on initial conditions and the Lemma above, 
we get for $t\in [z_j^i, z_{j+1}^i]$ that 
$$
\begin{array}{ll}
\|f(z(t))-f(l_{ij}(t)\| & \leq \|f(z(t))-f(\bar{z}_i(t))\|
+ \|f(\bar{z}_i(t)) - f(l_{ij}(t)\| \\
&\leq L\|z(t) - \bar{z}_i(t)\| + L\| \bar{z}_i(t) - l_{ij}(t)\| \\
&\leq L\exp(Lh)\|e_i\| + LM h\, \bar{h}^{r+\rho}
\end{array}
$$
for $Lh\leq \ln 2$. Inequality (\ref{p17}) 
together with (\ref{p14}) yield now that 
the inequalities 
\be
\begin{array}{lll}
\|e_{i+1}\| \leq \|e_i\|\, (1+hL\exp(hL))
&+& LM h^2\, \bar{h}^{r+\rho} \\
&+& h\, \bar{h}^{r+\rho} \left( \biggl{\|}  \frac{1}{m} 
\sum\limits_{j=0}^{m-1} \left(\int\limits_{0}^{1} g_{ij}(u) \, du
- \frac{1}{N} \sum\limits_{k=0}^{N-1} g_{ij}(u_k) \right) \biggr{\|}
+\e_1 \right)
\end{array}
\label{p18}
\ee
hold for $i=0,1,\ldots, n-1$ with probability at least $1-\delta$.
We now take into account the error of the mid-point rule, and solve 
the resulting
difference inequality with $e_0=0$. 
With probability at least $1-\delta$, we get that  
\be
\|e_i\| \leq C( h  +  1/N + \e_1)\, \bar{h}^{r+\rho} \, , 
\;\; i=0,1,\ldots, n,
\label{p19}
\ee
for a constant $C$ depending only on the parameters of the class $F^{r,\rho}$.
The total cost of computing 
$y_0,y_1,\ldots, y_n$ is equal in its deterministic part to
$c nm$ evaluations of partial derivatives of $f$. 
The~non-deterministic part includes 
\f
$O\left( n (\log\, n +\log\, 1/\delta) 
\min \{mN, 1/\e_1\} \right)$ quantum queries in the quantum setting, and 
\f
$O\left( n (\log\, n +\log\, 1/\delta) 
\min \{mN, (1/\e_1)^2 \} \right)$ evaluations of $f$ 
in the randomized setting.
\f
It follows from (\ref{p19}) with $N\geq n$ and $\e_1 = 1/n$ that
\be
\|e_i\| \leq C h\, \bar{h}^{r+\rho} \, ,
\label{p19a}
\ee
for $i=0,1,\ldots, n$, with probability at least $1-\delta$ 
(and a different constant $C$). 
Passing to~the approximation over $[a,b]$, 
we get for $t\in [x_i,x_{i+1}]$ the inequality
$$
\|z(t)-l(t)\| \leq \|z(t)-\bar{z}_i(t)\| + \|\bar{z}_i(t) -l_i(t)\|
\leq \exp(hL) \|e_i\| + M\, h\,\bar{h} ^{r+\rho} \, .
$$
This yields that  with probability at least $1-\delta$,  the error bound
\be
\sup_{t\in [a,b]} \|z(t)- l(t)\| \leq \tilde{C} \, h\, \bar{h}^{r+\rho} 
\label{p20}
\ee
holds, with the constant $\tilde{C}$ depending only on the parameters of 
the class $F^{r,\rho}$.
\f
Consider the quantum case. Neglecting for a while the logarithmic factors,
we have that error $O(1/(n(nm)^{r+\rho}))$ is achieved with cost
$O( nm+n^2)$. It is easy to see that the best choice in this case is
$m=n$. With a total number of $k$ quantum queries  and deterministic 
evaluations, we then achieve the error bound 
\be
\sup_{t\in [a,b]} \|z(t)- l(t)\| \leq C_1\, k ^{-(r+\rho+1/2)} \, ,
\label{p21}
\ee
with probability at least $1-\delta$.
This holds for all $f\in F^{r,\rho}$, and a constant $C_1$ depending only on 
the parameters of the class $F^{r,\rho}$. 
Hence, to compute an $\e$-approximation $l$ such that
$$\sup_{t\in [a,b]} \|z(t)- l(t)\| \leq \e$$
with probability at least $1-\delta$ for each $f\in F^{r,\rho}$, 
the algorithm uses    
$$O\left( (\log 1/\e + \log 1/\delta) \, (1/\e)^{1/(r+\rho+1/2)}\right)$$
quantum queries and deterministic evaluations (the logarithmic factors 
are again taken into account). This completes the proof 
of Theorem 2 in the quantum case.
\vsn
In the randomized setting, we proceed in a similar way, with $N\geq n^2$
and $m=n^2$.  With $k$ calls of $f$ or its partial derivatives 
(the logarithmic factors are for a while neglected), we get 
the error bound 
\be
\sup_{t\in [a,b]} \|z(t)- l(t)\| \leq C_2\, k^{-(r+\rho+1/3)}  \, .
\label{p22}
\ee
This holds with probability at least $1-\delta$ and a constant $C_2$ 
depending, as above, only on $F^{r,\rho}$.
Denote the left-hand side random variable in (\ref{p22}) 
by $X^{\omega}$, and the right-hand side by $h(k)$. We note that
$$\E (X^\omega)^2 =\int\limits_{ X^{\omega} > h(k) } 
(X^{\omega})^2 \, d \P(\omega)
+  \int\limits_{ X^{\omega} \leq h(k)} 
(X^{\omega})^2 \, d \P(\omega) 
\leq K^2 \delta + h(k)^2 $$
for all $f\in F^{r,\rho}$, where $K$ is a positive constant, depending only
on the parameters of the class $F^{r,\rho}$, such that 
$X^{\omega} \leq K$. To see that such a constant exists, note that 
the random variable $A_i(f)$ in (\ref{p12}) can be assumed bounded by 
$\|A_i(f)\| \leq 2M$, where $M$ is a bound on $\|g_{ij}\|$
(otherwise  $A_i(f)=0$ would be a better approximation). 
Proceeding  from (\ref{p18}) to (\ref{p19}) with $\e_1=3M$, 
we see from (\ref{p19}) that $X^{\omega}$ is bounded 
(in the deterministic sense) by
$\tilde{C}  \bar{h}^{r+\rho}$. Hence, the constant $K$ indeed exists.
\f
Take now $k$ to be the~minimal number such that $h(k) \leq \e/2$, so that
$k \asymp (1/\e)^{1/(r+\rho +1/3)}\, $, and set 
$\delta = \min \{\, 1/2, 3\e^2/(4K^2)\, \}$.
Then 
$\E \left( \sup_{t\in [a,b]} \|z(t)- l(t)\|\right)^2 \leq \e^2$ 
for all $f\in F^{r,\rho}$, 
which is achieved with cost $O\left( \log ( 1/\e) 
\cdot (1/\e)^{1/(r+\rho +1/3)} \right) $. 
This proves Theorem 2 in the randomized setting. \qed
\vsn
The upper bounds obtained in Theorem~2 are better than those 
from Theorem~1 for all $r$ and $\rho$.
For instance, for $r=0$ and $\rho=1$, if we neglect the logarithmic factors, 
Theorem 1 gives the bound
$O((1/\e)^{5/6})$ in the randomized setting, and $O((1/\e)^{3/4})$ 
in the quantum setting. In~Theorem~2 the respective bounds are
$O((1/\e)^{3/4})$ and  $O((1/\e)^{2/3})$. 
Nevertheless, we see from lower bounds in Theorem~1 that 
the gap still remains between the upper and lower bounds. 
\vsn
{\bf Remark 1} 
\f
We comment on the proof of Theorem 2, and  show a relation 
to Theorem 1. 
Looking at (\ref{p7}) we observe that, before starting randomized or
quantum computations, we can separate the~main part 
of $\int\limits_{0}^{1} g_{ij} (u)\, du$ by replacing this integral 
with $\int\limits_{0}^{1} s_{ij} (u)\, du +
\int\limits_{0}^{1} (g_{ij} (u) - s_{ij}(u))\, du $, where 
$s_{ij}$ is an approximation to $g_{ij}$. Using $l$ evaluations 
of $g_{ij}$ ($l\geq 1$), we can define $s_{ij}$ to have the~error of 
order $l^{-(r+\rho)}$, with the cost of one evaluation of $s_{ij}$
independent of $l$. We can next use randomized or quantum algorithms 
to compute  $\int\limits_{0}^{1} (g_{ij} (u) - s_{ij}(u))\, du $. 
In this way, we get  errors $\|e_i\|$ of order 
$$
(nm)^{-(r+\rho)} \left( n^{-1} + N^{-1} + \e_1 l^{-(r+\rho)}\right)
$$
with cost (up to logarithmic factors)
\be
nm+nml+ n \min\{mN, (1/\e_1)^\kappa \},
\label{p23}
\ee
where $\kappa=2$ in the randomized setting, and $\kappa=1$ on 
a quantum computer. By selecting optimal parameters, we get 
that the minimal (upper bound on the) error achieved with cost $k\;$
is equal to $k^{-(r+\rho +1/3)}$ in the randomized setting, and 
$k^{-(r+\rho +1/2)}$ on a quantum computer. 
Hence, by admitting $l\geq 1$ and by allowing a selection of $s_{ij}$ 
we do not arrive at better bounds than those given in Theorem 2, 
in which the functions $s_{ij} =0$ have simply been taken. 
\f
The upper bounds from Theorem 1 are a special case of (\ref{p23}), 
and can be obtained for sufficiently large $N$ by setting $m=1$, 
$\e_1 = n^{-1/(2r+2\rho +1)}$ and $l=n^{2/(2r+2\rho +1)}$
in the randomized setting, and 
$m=1$, $\e_1 = n^{-1/(r+\rho +1)}$ and $l=n^{1/(r+\rho+1)}$ on 
a quantum computer. 
\f
{\Large\section{ Scalar Autonomous Problems }}
\noindent
In this section, we study the solution of a scalar autonomous problem. The aim
is to compute the value of the solution at the end point of the interval
of integration. We give essentially tight upper and lower bounds on 
the complexity of this problem. 
In our previous paper \cite{rand5}, no lower bounds for this
problem were obtained. Upper bounds were discussed together with 
the general problem, which led to weaker results. 
\f
Note that the complexity of approximating the solution at only
one single point may differ from that of  
approximating the solution over the whole interval of integration, 
which is the subject of the proceding part of this paper. 
In particular, upper bounds for the former problem 
need not be valid for the latter one. 
\f
Consider problem (\ref{1}) with $d=1$, and the right-hand 
side function $f$ belonging to the class
\be
f\in \hat{F}^{r,\rho} = F^{r,\rho}\cap \{f:\;\; |f(y)| 
\geq p, \;\; y\in \rr\,\}\, ,
\label{p24}
\ee
for some $p>0$. Our aim is to compute the value $z(b)$ with accuracy $\e$
by randomized or quantum algorithms. Since
$$
t-a =\int\limits_\eta ^{z(t)} \frac{1}{f(s)} \, ds \, ,
$$
we equivalently look for the solution $y^*=z(b)$ of the nonlinear 
equation $H(y)=0$, where
\be
H(y)=\int\limits_\eta ^y \frac{1}{f(s)}\, ds - (b-a) \, .
\label{p25}
\ee
(The idea of transforming a scalar autonomous problem into a nonlinear 
equation was exploited, for example, in \cite{Brent} to derive a class
of nonlinear Runge-Kutta methods.)
\f
Note that 
$$
\frac{1}{D_0} |y-\bar{y}|\leq |H(y)-H(\bar{y})|\leq \frac{1}{p} 
|y-\bar{y}| \, ,
$$
for all $y$ and $\bar{y}$. 
\f
Given $y$, the computation of $H(y)$ reduces to the computation of the 
integral. Suppose that we have at our disposal a randomized or quantum 
algorithm for computing integrals, which
computes a random approximation $A(y)$ to $H(y)$ such that
\be
|H(y)-A(y)|\leq \e_1
\label{p26}
\ee
for some (small) $\e_1>0$,  with probability at least $3/4$, for any $f$ 
and $y$.
We denote the cost of this algorithm (dependent on a current setting)
 by $c(\e_1)$. 
\f
We now define algorithms for computing an approximation to 
$y^*=z(b)$ with error at most $\e$  with probability at least 
$1-\delta$, for all $f\in \hat{F}^{r,\rho}$.
We shall use the bisection method based on the~values $A(y)$. 
To get success probability at least $1-\delta$, we shall need 
inequality (\ref{p26}) to hold
with probability higher than $3/4$. Let $i^*$ be the minimal index $i$
for which $D_0(b-a)/(p2^{i+1}) \leq \e_1$, i.e., 
$i^*+1=\lceil \log\, \left( D_0(b-a)/(p\e_1) \right) \rceil$.
We need (\ref{p26}) to hold with probability at least $1-\delta_1$,
where $\delta_1= 1- (1-\delta)^{1/(i^* +1)}$. 
To increase the success probability in computing $A(y)$ from 
$3/4$ to $1-\delta_1$, we proceed 
in a standard way by computing the median of $k$ repetitions of 
the algorithm, where
\be
k=O(\log \, 1/\delta_1)=O( \log\, (i^* +1) +\log\, 1/\delta) = 
O(\log\log \, 1/\e_1 \,+\, \log\, 1/\delta)\; .
\label{p29}
\ee
Assume that $f(y)>0$ (the case $f(y)<0$ is analogous).
We start the bisection method from the~interval $[\alpha_0,\beta_0]=
[\eta, \eta+D_0(b-a)]$ containing $y^*$, and we set
$y_1=(\alpha_0+\beta_0)/2$. 
Given $[\alpha_i,\beta_i]$, we set $y_{i+1}=(\alpha_i+\beta_i)/2$ and 
select the next interval $[\alpha_{i+1},\beta_{i+1}]$  based
on the sign of $A(y_{i+1})$. We stop the iteration at first index $i$, 
call it $i^{{\rm bis}}$, for which $|A(y_i)|\leq 2\e_1$ (we shall discuss
this termination criterion and the correctness of the selection of
successive intervals in a while).  
\f
Note that for any $j\leq i^*$,  inequalities
\be
|H(y_{i+1}) - A(y_{i+1})|\leq \e_1 \;\;\;\;\; i=0,1,\ldots, j
\label{n1}
\ee
hold (simultaneously) with probability 
at least $(1-\delta_1)^{i^*+1} = 1-\delta$. 
\f
Assume that (\ref{n1}) is satisfied. We show that the number
of bisection steps satisfies $i^{{\rm bis}} \leq i^*+1$. 
Suppose that the termination condition is not fulfilled by the
$i^*$-th step, i.e., $|A(y_i)| > 2\e_1$ for $i=1,2,\ldots, i^*$.
Then the selection of the interval 
$[\alpha_{i^*},\beta_{i^*}]$ made on the basis of $A(y_{i^*})$,
as well as the selection of all proceeding intervals, is correct.
(In fact, it suffices for this that $|A(y_i)| > \e_1$, since 
the signs of $A(y_{i})$ and  $H(y_{i})$ are then the same by (\ref{n1}).\,)
Hence, we have
$$
|y_{i^*+1}-y^*| \leq |\alpha_{i^*} - \beta_{i^*}|/2 = 
|\alpha_0 -\beta_0|/2^{i^*+1},
$$
and
$$
|A(y_{i^*+1})| \leq |H(y_{i^*+1})|+|A(y_{i^*+1}) - H(y_{i^*+1})|\leq 
|y_{i^*+1} -y^*|/p + \e_1 \leq |\alpha_0 -\beta_0|/(p2^{i^*+1}) + \e_1
\leq 2\e_1.
$$
Since the termination condition is now satisfied,  we have in this case that
$i^{{\rm bis}} = i^*+1$. In any case,
the desired bound on $i^{{\rm bis}}$ holds, as claimed. 
In terms of $\e_1$, we have that
\be
i^{{\rm bis}} \leq \lceil \log\left( D_0(b-a)/(p\e_1) \right) \rceil
\; .
\label{p27}
\ee
Take now $\e_1=\e / (3D_0)$. 
Then, terminating after $i^{{\rm bis}}$ steps, 
we arrive at the $\e$-approximation $y_{i^{{\rm bis}}}$ to $y^*$, 
since
\be
|y_{i^{{\rm bis}}}-y^*| \leq D_0 |H(y_{i^{{\rm bis}}})| \leq 
D_0( |A(y_{i^{{\rm bis}}})| + |H(y_{i^{{\rm bis}}}) - A(y_{i^{{\rm bis}}})|) 
\leq 3D_0 \e_1 =\e \; .
\label{p28}
\ee
As in the case of (\ref{n1}), this holds with probability at least $1-\delta$.
\f
Summarizing, the described algorithm returns the approximation 
(random variable) $y_{i^{{\rm bis}}}$ such that the bound 
$|y_{i^{{\rm bis}}}-z(b)|\leq \e$ holds with probability at least $1-\delta$, 
with total cost 
\be
c(\e_1)\, k\, i^{{\rm bis}}=
O\left( c(\e_1)\,  \left( \log\log\, 1/\e +
\log\,1/\delta \right) \log\, 1/\e \right) .
\label{p291}
\ee
Using known results on integration, we now estimate $c(\e_1)$. 
Since $f\in \hat{F}^{r,\rho}$, the function $1/f(s)$ is in the 
H\"older class $F^{r,\rho}$ (over the finite interval 
$s\in [\eta, \eta +D_0(b-a)]$)
with certain parameters $\tilde{D}_0, \tilde{D}_1,\ldots,
\tilde{D}_r, \tilde{H}$ depending on $D_0,D_1,\ldots, D_r, H$ and $p$.
\f
Consider the quantum setting. There exists an algorithm for 
computing  integrals of $1/f(s)$ with cost 
$c(\e_1)=O\left((1/\e_1)^{1/(r+\rho+1)} \right)$ quantum queries,
see \cite{Novak}. This leads to the following 
complexity bound for our problem 
\be
\comp^{{\rm quant}} (\hat{F}^{r,\rho}, \e, \delta) = 
O\left(\left( 1/\e\right) ^{ 1/(r+\rho+1) }\, 
\left(\log\log \, 1/\e + \log \, 1/\delta \right) \; \log \, 1/\e \right)  \, . 
\label{p32}
\ee
{\bf Remark 2}
\f
To establish the cost of an algorithm in the quantum setting,
we have to count the number of applications of a quantum query 
operator $Q_f$ for $f$.
Calculating the cost $c(\e_1)$ above we have taken into account 
the number of queries $\tilde{Q}_{1/f}$ for $1/f$.
However, a query for $1/f$ for $f\in \hat{F}^{r,\rho}$ can be simulated 
by a query for $f$ (and vice versa), see Lemma 4 in \cite{Heinrich}.
Hence, the upper bound in terms of both units remains the same. 
\vsn
Consider the randomized setting. There exists an algorithm  
approximating integrals with 
the~mean square error (\ref{3a}) bounded by $\e_1/2 $, 
and cost
$c(\e_1)  =O\left((1/\e_1)^{1/(r+\rho+1/2)} \right)$ evaluations of $f$.
We use it to compute an approximation $A(y)$ to $H(y)$, for a given $y$.
By the Markov inequality, error bound (\ref{p26}) holds for $A(y)$ with
probability at least $3/4$.
We now follow the~steps between relations (\ref{p26}) and (\ref{p291}) 
above to get the approximation $y_{i^{{\rm bis}}}$ to $z(b)$ such that
\be
X^{\omega} := |y_{i^{{\rm bis}}}-z(b)|\leq \e 
\label{p321}
\ee
 with probability at least $1-\delta$, for all $f$. 
By (\ref{p291}), the cost of computing $y_{i^{{\rm bis}}}$
is  
$$
O\left(  (1/\e)^{1/(r+\rho+1/2)} \,
 \left( \log\log\, 1/\e + \log\,1/\delta \right)
\log\, 1/\e\, \right).
$$ 
To estimate the mean square error of $y_{i^{{\rm bis}}}$, we proceed 
in a similar way as we did
in the final part of the proof of Theorem~2. We 
replace $\e$ in (\ref{p321}) by $\e/2$, which influences the cost
only by a constant factor, and we write
$$\E (X^\omega)^2 =\int\limits_{ X^{\omega} > \e/2 } 
(X^{\omega})^2 \, d \P(\omega)
+  \int\limits_{ X^{\omega} \leq \e/2} 
(X^{\omega})^2 \, d \P(\omega) 
\leq K^2 \delta + \e^2/4 $$
for all $f\in \hat{F}^{r,\rho}$. Here, $K$ is a positive constant 
depending only
on the parameters of the class $\hat{F}^{r,\rho}$ such that 
$X^{\omega} \leq K$.
The choice $\delta=3\e^2/(4K^2)$ gives the bound
$$
\sup_{f\in \hat{F}^{r,\rho}} (\E (X^\omega)^2)^{1/2} \leq \e\; ,
$$
which is achieved with cost 
$$O\left(  (1/\e)^{1/(r+\rho+1/2)} \,
 \left( \log\log\, 1/\e + \log\,1/\e\right) 
\log\, 1/\e\, \right)=
O\left(  (1/\e)^{1/(r+\rho+1/2)} \,
 \left( \log\,1/\e\right)^2 \right).
$$
\f
This yields an upper bound 
\be
\comp^{{\rm rand}} (\hat{F}^{r,\rho}, \e) = 
O\left(  (1/\e)^{1/(r+\rho+1/2)} \,
 \left( \log\,1/\e \, \right)^2 \right) 
\label{p31}
\ee
on the complexity. 
Hence, up to logarithmic factors, we are able to solve our problem 
 at cost of one single integration. 
\vsn
We now turn to lower bounds on the randomized and quantum complexity.
Let $\phi$ be any algorithm based on evaluations 
of $f$ or its derivatives at possibly random points in the randomized setting,
and on quantum queries for $f$ and classical evaluations 
of $f$ or its derivatives in the~quantum setting. 
Assume that $\phi$ computes an approximation to $z(b)$ with
error at most $\e$, for any scalar problem (\ref{1}) 
with $f\in \hat{F}^{r,\rho}$.
We estimate from below the number of evaluations (queries) used by 
$\phi$, by reducing the problem  to the summation of real numbers.
\f
Without loss of generality, let $[a,b]=[0,1]$. For $n\geq 1$, let
$\lambda_0,\lambda_1, \ldots,\lambda_{n-1}$ be numbers of at most unit
absolute value, and define the function $g(y)=1/f(y)$
as  follows.
\f
Consider the uniform partition of $[\eta, \eta+1/2]$ with points $y_i=
\eta+i/(2n)$ for $i=0,1,\ldots,n$. We let $h_i\in F^{r,\rho}$, 
where $i=0,1,\ldots, n-1$,
 be functions with the following properties:
$$\mbox{ $h_i$ has support $[y_i,y_{i+1}]$ },\;\;\; h_i(y)\geq 0, $$
$$ \max h_i(y) = h_i((y_i+y_{i+1})/2) = c_1(y_{i+1} - y_i)^{r+\rho},$$
$$\int\limits_{y_i}^{y_{i+1}} h_i(y)\, dy = c_2 (y_{i+1} - y_i)^{r+\rho+1},$$
where $c_1$ and $c_2$ are known positive constants depending only on the 
parameters of the class $F^{r,\rho} \;$ (and not on $i$ and $n$). 
Such functions are 
often used in proving lower bounds and their construction is well known.
\f
We define $g(y)= 1+ \sum\limits_{i=0}^{n-1} \, \lambda_ih_i(y)$. Then
$g\in \hat{F}^{r,\rho}$ for sufficiently large $n$, and the same 
holds for $f$  (with different constants).
Since  $3/4\leq f(y)\leq 3/2$ for sufficiently large $n$, we have that
$\eta+3/4 \leq z(1)\leq \eta + 3/2$, and we can write 
\be
1=\int\limits_{\eta}^{z(1)} 1/f(y)\, dy = \int\limits_{\eta}^{\eta+1/2} 
1/f(y)\, dy + 
\int\limits_{\eta+1/2}^{z(1)} 1/f(y)\, dy = c_3 n^{-(r+\rho+1)}
\sum\limits_{i=0}^{n-1} \lambda_i + z(1) -\eta\, .
\label{p33}
\ee
This yields that
\be
\frac{1}{n}\sum\limits_{i=0}^{n-1} \lambda_i = 
\frac{1-z(1)+\eta}{c_3\, n^{-(r+\rho)}},
\label{p34}
\ee
where $c_3$ is a known positive constant.
\f
Consider first the quantum setting. Let $\zeta$ 
(a random variable) be an $\e$-approximation to $z(1)$
 computed by the algorithm $\phi$ for the right-hand side 
$f$ defined above. We have that
$$
|z(1)-\zeta| \leq \e
$$
with probability at least $3/4$ (we take $\delta\leq 1/4$). Hence,
$$
\zeta_1 = \frac{1-\zeta+\eta}{c_3\, n^{-(r+\rho)}} \;\;\;
\mbox{ is an approximation to } \;\;\;
S= \frac{1}{n}\sum\limits_{i=0}^{n-1} \lambda_i 
$$
with error at most $\e_2 = \e\, n^{r+\rho}/c_3 \,$ and probability 
at least $3/4$. The lower bound of Nayak and Wu,
see (\ref{7d}), gives 
that the number of queries for $\Lambda=[\lambda_0,\lambda_1, \ldots,
\lambda_{n-1}]$ must be at least 
$\Omega(\min\{n, 1/\e_2\})$. This is also a lower bound on the number of
queries for $1/f$ (and for $f$) needed in the algorithm $\phi$.
We now take $n\asymp (1/\e)^{1/(r+\rho +1)}$, and we conclude that
$$
\comp^{{\rm quant}} (\hat{F}^{r,\rho}, \e, 1/4) = 
\Omega \left( (1/\e)^{1/(r+\rho +1)} \right) \, .
$$
In the randomized setting, let
$
(\E|z(1)-\zeta|^2)^{1/2} \leq \e\, .
$
Then 
$
(\E|S-\zeta_1|^2)^{1/2} \leq \e\, n^{r+\rho} / c_3 =\e_2
$
and the same inequality holds for $\E|S- \zeta_1|$. Due to (\ref{7c}), 
the number of accesses to $\lambda_i$ must be at least
$\Omega\left( \min\{n, (1/\e_2)^2 \} \right)$. 
This is also a lower bound on the number of the number of evaluations 
of $f$ or its derivatives. We now take 
$n\asymp (1/\e)^{1/(r+\rho +1/2)}$ to get 
$$
\comp^{{\rm rand}} (\hat{F}^{r,\rho}, \e) = 
\Omega \left( (1/\e)^{1/(r+\rho +1/2)} \right) \, .
$$
\noindent
We have shown the following
\vsn
{\bf Theorem 3} $\;\;$ {\it Consider 
the scalar autonomous problem described in the beginning of this section, 
with a right-hand side $f$ in 
the class $\hat{F}^{r,\rho}$. There exist positive constants
$P_i$ ($i=3,4,5,6$) depending only on the parameters of the class 
$\hat{F}^{r,\rho}$
such that, for sufficiently small $\e$ and $\delta$, the~following 
complexity bounds hold true. 
\f
In the randomized setting
\be
\comp^{{\rm rand}} (\hat{F}^{r,\rho}, \e) \leq
P_3 \left( \left( \frac{1}{\e}\right)^{1/(r+\rho+1/2)} \,
 \left( \log \frac{1}{\e}\right)^2 \right)
\label{th31}
\ee
and 
\be
\comp^{{\rm rand}} (\hat{F}^{r,\rho}, \e) \geq
P_4 \left(\frac{1}{\e} \right)^{1/(r+\rho +1/2)}  \, .
\label{th32}
\ee
\f
In the quantum setting
\be
\comp^{{\rm quant}} (\hat{F}^{r,\rho}, \e, \delta) \leq 
P_5 \left(\left( \frac{1}{\e}\right)^{1/(r+\rho+1)} 
\left( \log\log \frac{1}{\e} + \log \frac{1}{\delta} \right)
\log \frac{1}{\e}\right) 
\label{th33}
\ee
and, for $0 < \delta\leq 1/4$, 
\be
\comp^{{\rm quant}} (\hat{F}^{r,\rho}, \e, \delta) \geq 
\comp^{{\rm quant}} (\hat{F}^{r,\rho}, \e, 1/4) \geq 
P_6 \left( \left(\frac{1}{\e}\right)^{1/(r+\rho +1)} \right) \, .
\label{th34}
\ee }
\qed
\f
Note that in  both randomized and quantum settings upper and lower bounds
in Theorem 3 are
matching, up to logarithmic factors. 
The question of finding matching upper and lower bounds for the general 
problem (\ref{1}) still remains open.
\vsn

\end{document}